\documentclass[reprint, amsmath,amssymb, aps]{revtex4-1} 
\usepackage[dvipdfmx]{graphicx} 
\usepackage{dcolumn}
\usepackage{bm}
\usepackage{color}

\begin{document}
\preprint{IPMU16-0025}

\title{Conformal Bootstrap Dashing Hopes of Emergent Symmetry}


\author{Yu Nakayama}
\author{Tomoki Ohtsuki}
\affiliation{Kavli Institute for the Physics and Mathematics of the Universe (WPI),  \\ University of Tokyo, 5-1-5 Kashiwanoha, Kashiwa, Chiba 277-8583, Japan} 

\date{\today}

\begin{abstract} 
We use the conformal bootstrap program to derive necessary conditions for emergent symmetry enhancement from discrete symmetry (e.g. $\mathbb{Z}_n$) to continuous symmetry (e.g. $U(1)$) under the renormalization group flow. In three dimensions, in order for $\mathbb{Z}_2$ symmetry to be enhanced to $U(1)$ symmetry, the conformal bootstrap program predicts that the scaling dimension of the order parameter field at the infrared conformal fixed point must satisfy $\Delta_1 > 1.08$.  We also obtain the similar conditions for $\mathbb{Z}_3$ symmetry with  $\Delta_{1} > 0.580$ and $\mathbb{Z}_4$ symmetry with $\Delta_1 > 0.504$ from the simultaneous conformal bootstrap analysis of multiple four-point functions.
Our necessary conditions impose severe constraints on many controversial physics such as the chiral phase transition in QCD,  the deconfinement criticality in N\'eel-VBS transitions and anisotropic deformations in critical $O(n)$ models.
In some cases, we find that the conformal bootstrap program dashes hopes of emergent symmetry enhancement proposed in the literature.

\end{abstract}
\pacs{Valid PACS appear here}
\maketitle 
\section{Introduction} 
Symmetry in nature is the most helpful guideline to understand physics.
Beauty of the symmetry is that it may not have a microscopic origin, but it
 may appear as an emergent phenomenon. Such emergent symmetry plays a significant role in theoretical physics. 

Take a lattice system for example. Suppose the defining Hamiltonian possesses certain discrete or continuous symmetry. This does not mean that the infrared (IR) physics has the same symmetry. Rather, it often shows enhanced symmetry, especially when the system is at criticality. Indeed, emergence of global continuous symmetry out of discrete lattice symmetry is ubiquitous in strongly interacting systems, and it has played a key role in understanding the nature of quantum criticality that is outside the scope of the traditional Wilson-Landau-Ginzburg (WLG) paradigm of phase transitions \cite{SVBS,2004PhRvB..70n4407S}.

In this paper, we derive universal necessary conditions for such emergent symmetry enhancement from discrete symmetry to continuous symmetry under the renormalization group (RG) flow  by using the recently developed technique of numerical conformal bootstrap program in three-dimensions \cite{ElShowk:2012ht,Kos:2013tga,El-Showk:2014dwa,Kos:2014bka,Nakayama:2014lva,Nakayama:2014sba,Simmons-Duffin:2015qma,Kos:2015mba,Iliesiu:2015qra}. We will show that the conformal symmetry imposes a strong constraint on when the emergent symmetry enhancement can or cannot occur.

Let us rephrase the question in terms of conformal field theories (CFTs).
 Suppose we have a system with emergent $U(1)$ symmetry in the IR. Can we realize the same system with smaller discrete symmetry (e.g. $\mathbb{Z}_2 \in U(1)$)  without fine-tuning? The $\mathbb{Z}_2$ symmetry forbids the perturbation of the $U(1)$ symmetric fixed point under the smallest charged operators that are $\mathbb{Z}_2$ odd. However, with the only $\mathbb{Z}_2$ symmetry, one cannot forbid a perturbation by twice $U(1)$ charged operators that are $\mathbb{Z}_2$ even. In order to obtain the emergent $U(1)$ symmetry, all the $\mathbb{Z}_2$ even but $U(1)$ charged operators must be irrelevant. The conformal bootstrap program tells exactly when this can happen.
In this case, we find that the scaling dimension of  the $\mathbb{Z}_2$ odd order parameter field must satisfy $\Delta_1 > 1.08$ in three dimensions. Otherwise, we always have $\mathbb{Z}_2$ even but $U(1)$ charged relevant deformations that we cannot forbid without fine-tuning.

Prior to our work, hopes of the emergent symmetries have relied on explicit ultraviolet Lagrangian or Hamiltonian with naive dimensional counting, or, at best, with the perturbative computations e.g. large $N$ expansions or $\epsilon$ expansions. Our necessary conditions from the conformal bootstrap program are non-perturbative, rigorous and universal, so they should be applied to any critical phenomena in nature as long as the conformal symmetry is realized at the fixed point.

In this paper, among many possibilities, we offer applications to two widely discussed controversies in the theoretical physics community. The one is finite temperature chiral phase transition in Quantum Chromo Dynamics (QCD) and the other is the deconfinement criticality in N\'eel-Valence Bond Solid (VBS) transitions. We also test our necessary conditions against anisotropic deformations of $O(n)$ critical vector models. In certain cases, we find that the conformal bootstrap program dashes hopes of emergent symmetry enhancement proposed in the literature.

\section{Necessary conditions for emergent symmetry enhancement from conformal bootstrap}\label{sec:2}
The foremost basis of our claim is the conformal hypothesis: under the RG flow, the system reaches a critical point described by
a unitary CFT. In particular, not only scale symmetry but also Lorentz and special conformal symmetry should emerge. The hypothesis seems to be valid in many classical as well as quantum critical systems as long as we trust the effective field theory description with emergent Lorentz symmetry. In particular, in the examples we will study in section \ref{sec:3}, there are no perturbative candidates for the Virial current in the effective action, which is the obstruction for conformal invariance in scale invariant field theory, so the scale invariance most likely implies conformal invariance. See e.g. \cite{Nakayama:2013is} for a review on this argument.

Once conformal invariance is assumed, we may study the consistency of four-point functions that results in the conformal bootstrap equations. In our case, we are interested in the consistency of four-point functions $\langle O_q O^\dagger_{q} O_{q'} O^\dagger_{q'} \rangle$ of $U(1)$ charge $q$ local scalar operators $O_q$, whose scaling dimension is denoted by $\Delta_q$, with the crossing equations and unitarity, whose idea was first developed in four dimensional CFTs in \cite{Rattazzi:2010yc}. 
By mapping the crossing equations in unitary CFTs to a semi-definite problem, numerical optimization yields a bound on the scaling dimension of the operators that appear in the operator product expansion (OPE) e.g. $O_q \times O_{q'} \sim O_{q+q'}$. See Appendix \ref{app:a} for the details of our implementation.

\begin{figure}[htbp]
  \includegraphics[width=7cm,clip]{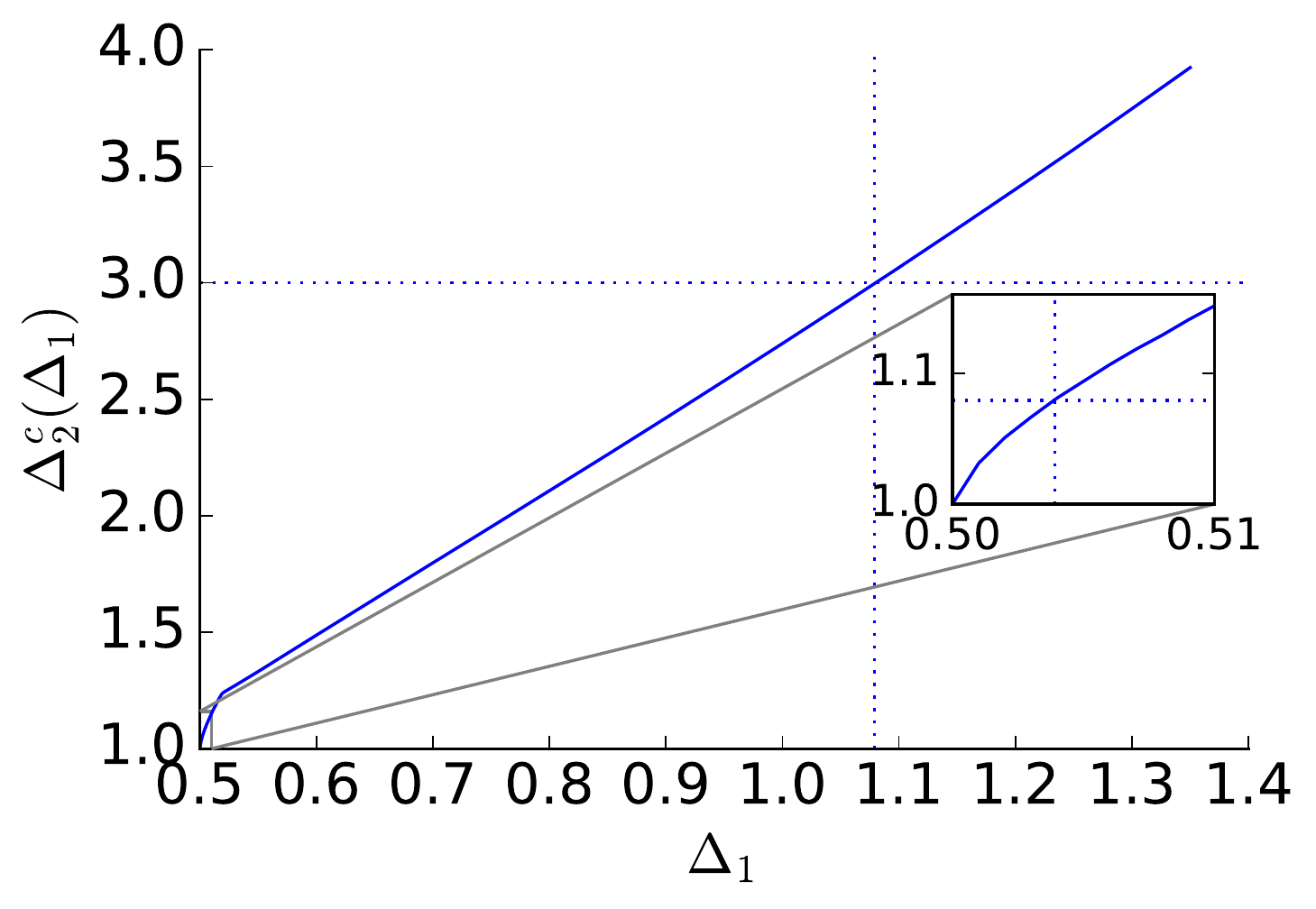}
  \caption{The upper bound on the scaling dimension $\Delta^c_2$ of the lowest dimensional charge two scalar operator appearing in  $O_1 \times O_1$ OPE  as a function of $\Delta _1 $. The same bound applies to $O_2\times O_2 \sim O_4$.}
  \label{fig:charge_2}
\end{figure}

Let us begin with emerging $U(1)$ symmetry from $\mathbb{Z}_2$. The upper bound on $\Delta_{2}$ as a function of $\Delta_{1}$ in $U(1)$ symmetric CFTs is straightforwardly obtained as in \cite{Kos:2014bka} by studying $\langle O_1 O_{1}^\dagger O_{1} O_{1}^\dagger \rangle$. The plot in Fig.\ref{fig:charge_2} shows the necessary condition  $\Delta_1 > 1.08$ for the symmetry enhancement as the bound when $\Delta_{2}$ can be larger than $3$, at which point $O_{2}$ may become irrelevant. In other words, when $\Delta_1 <1.08$, $O_{2}$ is always relevant and symmetry enhancement does not occur.

For the $\mathbb{Z}_3$ enhancement, we study the simultaneous consistency of three four-point functions $\langle O_1 O_{1}^\dagger O_{1} O_{1}^\dagger \rangle$,  $\langle O_1 O_{1}^\dagger O_{2} O_{2}^\dagger \rangle$  and $\langle O_{2} O_{2}^\dagger O_{2} O_{2}^\dagger \rangle$ from the mixed correlator conformal bootstrap analysis \cite{Kos:2015mba,Lemos:2015awa}. In order to make the bound relevant for us, we make two additional assumptions: (1)  all the charge four operators are irrelevant (2) all the charge neutral operators (above the identity) have scaling dimension larger than $1.044$. The latter assumption is motivated from our setup because it is easy to numerically prove it by using the conformal bootstrap analysis that if there exists a neutral scalar operator with scaling dimension less than $1.044$, there also exists another neutral scalar operator whose scaling dimension is less than 3 (see Appendix \ref{app:2.5} for details). However, in all of our applications, there is only one neutral scalar operator that must be tuned, so the assumption is justifiable.
\begin{figure}[htbp]
	\begin{center}
  \includegraphics[width=9.0cm,clip]{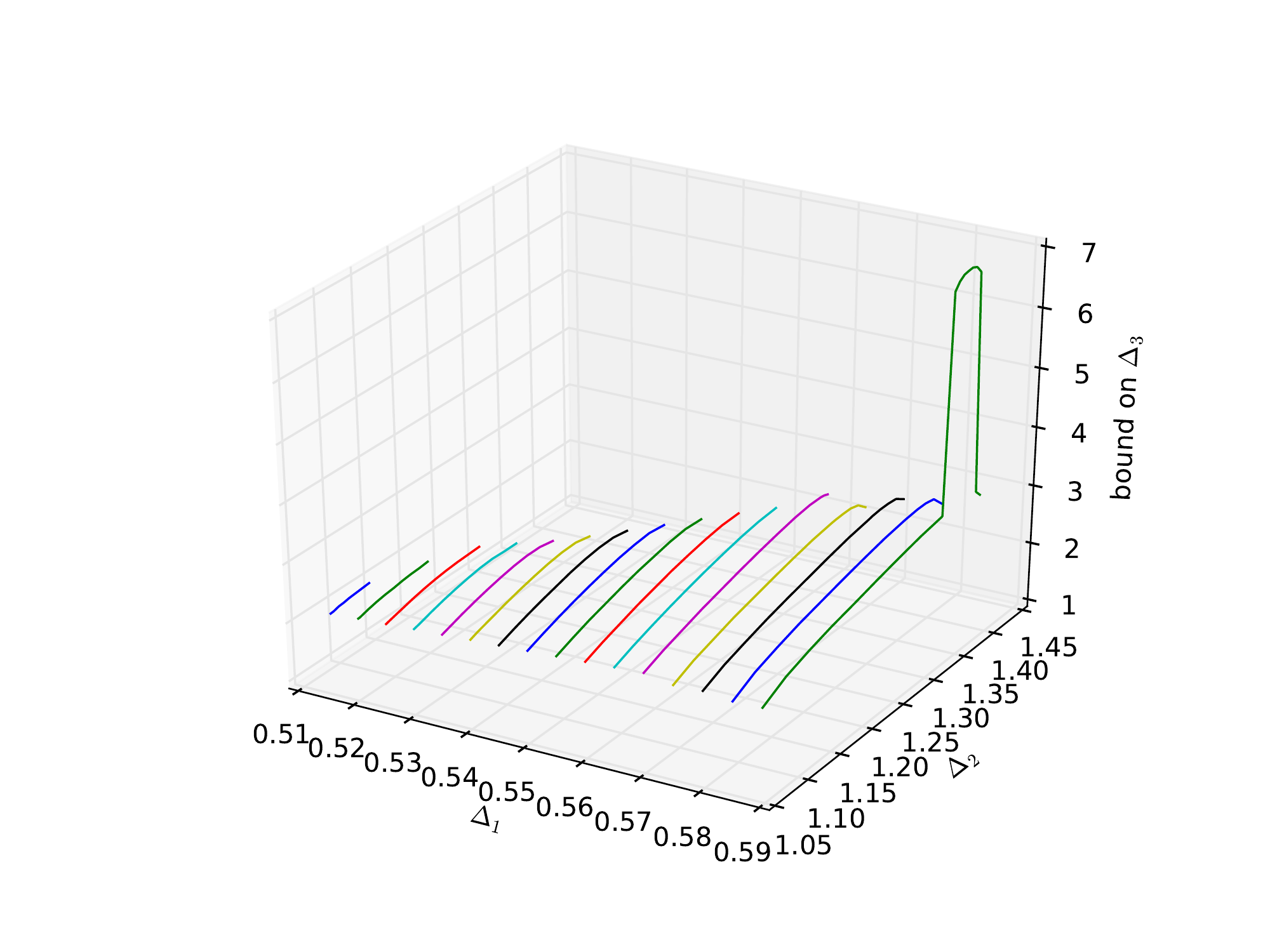}
  \end{center}
  \caption{Upper Bounds on the scaling dimension of the lowest dimensional charge three scalar operator appearing in $O_1 \times O_2$ OPE as a function of $\Delta _1 $ and $\Delta _2$. The jump in the bounds appears as soon as they touch the value 3. Note that $1.08<\Delta_2 < \Delta^c_2 (\Delta_1)$ must hold from the assumption that all the charge four operators are irrelevant and the bound in Fig.\ref{fig:charge_2}.}
  \label{fig:charge_3}
\end{figure}

Fig.\ref{fig:charge_3} shows the bound on $\Delta_{3}$ as a function of $\Delta_1$ and $\Delta_{2}$. When $\Delta_1 \ge 0.585$, there exists an allowed region of $\Delta_2$ where $\Delta _ 3$ can be irrelevant. As soon as the bound on $\Delta_{3}$ touches $3$, it shows a conspicuous jump that is similar to the one observed in the fermionic conformal bootstrap analysis \cite{Iliesiu:2015qra}. Without knowing the value of $\Delta_{2}$, the plot shows that the necessary condition is $\Delta_1 > 0.580$. See Appendix \ref{app:3} for two-dimensional projections of the plot.

In the similar manner, we can study the bound on $\Delta_{4}$ for the $\mathbb{Z}_4$ enhancement. We obtain the simplest bound by studying $\langle O_1 O_{1}^\dagger O_{1} O_{1}^\dagger \rangle$ and $\langle O_{2} O_{2}^\dagger O_{2} O_{2}^\dagger \rangle$ independently, which immediately gives $\Delta_1 > 0.504$ (see Fig.\ref{fig:charge_2}). The study of the simultaneous consistency of three four-point functions $\langle O_1 O_{1}^\dagger O_{1} O_{1}^\dagger \rangle$,  $\langle O_1 O_{1}^\dagger O_{2} O_{2}^\dagger \rangle$  and $\langle O_{2} O^\dagger_{2} O_{2} O_{2}^\dagger \rangle$ gives a stronger bound in principle, but in practice, without introducing further assumptions, it does not improve much.

\section{Applications}\label{sec:3}
\subsection{Chiral phase transition in QCD}
The order of chiral phase transition in finite temperature  QCD has been controversial over many years without reaching a consensus. In the WLG paradigm, we may translate the problem into (non-)existence of RG fixed point in a certain three-dimensional WLG model whose order parameter is given by the quark bilinear ``meson" field $\Phi_{\bar{i}j} = \bar{\psi}_{\bar{i}} \psi_j$ (where $\bar{i},j = 1,2$ runs the number of ``massless quarks" in nature). To reveal the nature of the RG flow, it is crucial to discuss whether the anomalous $U(1)_A$ symmetry is restored in the IR limit of the effective WLG model. If the $U(1)_A$ symmetry is restored, we expect that the chiral phase transition is described by an RG fixed point with the symmetry of $ O(4) \times U(1)_A$ \cite{Pelissetto:2013hqa,Nakayama:2014sba}.  Otherwise, it is described by RG flow only with the symmetry of $O(4)$  \cite{Pisarski:1983ms}. Here $O(4) \sim SU(2)_R \times SU(2)_L$ is the non-anomalous flavor symmetry of two massless quarks.

In \cite{Aoki:2012yj,Aoki:2013zfa,Kanazawa:2015xna}, it was shown that under mild assumptions, $\mathbb{Z}_2$ subgroup of the anomalous $U(1)_A$ is microscopically restored, which raises the second question if the $\mathbb{Z}_2$ can be further enhanced to the full $U(1)_A$ under the RG flow of the effective WLG model in three dimensions. This is exactly the problem we have discussed in section \ref{sec:2}, and the conformal bootstrap program gives a definite answer.

A study of the RG properties of this effective WLG model is notoriously hard, but the conformal bootstrap analysis of \cite{Nakayama:2014sba} tells that the scaling dimension of the $\mathbb{Z}_2$ odd operator at the  $O(4) \times U(1)_A$ symmetric fixed point is $\Delta_1 = 0.82(2)$ (see Appendix \ref{app:b} for more details). 
It turns out that this value does not satisfy the necessary condition for the $U(1)$ symmetry enhancement that we have derived in section \ref{sec:2}. We therefore conclude that the microscopic $O(4) \times \mathbb{Z}_2$ symmetry cannot be enhanced to $O(4) \times U(1)_A$ without fine-tuning. It means that the chiral phase transition in QCD does not accompany the full restoration of the $U(1)_A$ symmetry and does not show the second order phase transition described by the fixed point studied in \cite{Pelissetto:2013hqa,Nakayama:2014sba} unless further symmetry enhancement is assumed.

\subsection{Deconfinement criticality in N\'eel-VBS transitions} 
The deconfinement criticality in N\'eel-VBS transitions in $2+1$ dimensions is proposed to be an example of critical phenomena whose description is beyond the traditional framework of WLG effective field theory. In \cite{SVBS,2004PhRvB..70n4407S}, they argued that the effective field theory description with $N$ component spin  near the critical point is given by the non-compact $CP^{N-1}$ model \cite{PhysRevLett.32.292,PhysRevLett.71.1911,PhysRevB.70.075104}, or a $U(1)$ gauge theory coupled with $N$ charged scalars with $SU(N)$ flavor symmetry. While there has been no rigorous proof, it was argued  that the system shows a conformal behavior once one can tune one parameter, the bare mass of the charged scalars. 

However, it turns out that the actual realization of this critical behavior in the lattice simulation has been controversial over years. From the effective field theory viewpoint, a difficulty comes from the existence of monopole operators. One can argue that the lattice symmetry forbids the smallest charged monopole operator, but not necessarily so for the higher charged monopole operators \cite{2013PhRvL.111m7202B,2013PhRvB..88v0408H}. For instance, if we use the rectangular lattice, one can only preserve the $\mathbb{Z}_2$ subgroup of the $U(1)$ monopole charge, and if we use the  honeycomb lattice it is $\mathbb{Z}_3$ and if we use the square lattice it is $\mathbb{Z}_4$. If the higher charged monopole operators that are not forbidden by the lattice symmetry are relevant, then we cannot reach the non-compact $CP^{N-1}$ model in the IR limit without further fine-tuning, and we typically expect the first order phase transition in lattice simulations.

Therefore the central question we should address is under which condition, the higher charged monopole operators become irrelevant once we know the scaling dimension of the lowest monopole charged operator. Again this is precisely the question we have studied in section \ref{sec:2}. 
To reiterate our results, in order to obtain $U(1)$ symmetry enhancement, we need $\Delta_{1} > 1.08$  from $\mathbb{Z}_2$, $\Delta_{1}>0.580$ from $\mathbb{Z}_3$ and $\Delta_{1} > 0.504$ from $\mathbb{Z}_4$. 

In the following, we critically review various predictions about the nature of N\'eel-VBS phase transitions in the literature for different $N$. 
We may find a convenient summary of the scaling dimensions of operators proposed in the literature in Appendix \ref{app:b}. 

Let us begin with the $N=2$ case, which has the most experimental significance. The predictions of $\Delta_{1}$ in the literature ranges between $0.57$ and $0.68$. For whichever values, our necessary condition for $\mathbb{Z}_4$ tells that the charge four operators can be irrelevant and it is consistent with the observation of the second order phase transition on the square lattice.
On the other hand, our necessary condition $\Delta_{1} >1.08$ for $\mathbb{Z}_2$ tells that the charge two monopole operator is relevant so we expect the first order phase transition on the rectangular lattice as observed indeed in \cite{2013PhRvL.111m7202B}. 

The most controversial question is if the charge three monopole operator is relevant or not. Our necessary condition $\Delta_{1}>0.580$ is consistent with that it is either relevant or irrelevant, depending on the value of $\Delta_1$. 
We note that the scaling dimensions obtained in \cite{2015arXiv150402278S} (i.e. $\Delta_1 = 0.579(8), \Delta_{2} = 1.42(7)$ and $\Delta_{3}= 2.80(3)$) are very close to the bound. In particular, our results show that, given  their values of $\Delta_1$ and $\Delta_2$, the charge three monopole operator must be relevant, supporting their claim that they observed the first order phase transition on the honeycomb lattice.

Let us next consider the $N=3$ case.  In \cite{2013PhRvB..88v0408H}, they obtained $\Delta_1 = 0.785$, so our necessary condition implies that it can show the second order phase transition on the square lattice but it cannot on the rectangular lattice, in agreement with what is observed. The direct measurement of $\Delta_{2} = 2.0 $  there turns out to be close to but slightly below our bound with $\Delta_1 = 0.785$. However, we also note that the earlier estimate of $\Delta_1 = 0.71(2)$ in \cite{2009PhRvB..80r0414L} may be inconsistent with $\Delta_{2} = 2.0 $.

For $N=4$, our result reveals inconsistency among the literature. In \cite{2013PhRvB..88v0408H} they obtained $\Delta_1 = 0.865$, and our condition predicts that the charge two monopole operator is relevant. On the other hand in \cite{2013PhRvL.111m7202B}, they claim that the charge two monopole operator is irrelevant and the phase transition is second order on the rectangular lattice. These statements cannot be mutually consistent, and one of them or our conformal hypotheses must be wrong.

For $N=5$, the situation is again subtle. \cite{2013PhRvL.111m7202B} claims that it shows the second order phase transition on the rectangular lattice. Our result then demands that $\Delta_{1} > 1.08$ to make charge two monopole operator irrelevant. The values they obtained on square and honeycomb lattices $\Delta_{1} =  1.0(1)$ are quite marginal. In contrast, the value of $\Delta_{1} = 0.85(1)$ they obtained on rectangular lattice is clearly inconsistent, so it is likely that the charge two monopole operator is actually relevant and the phase transition on the rectangular lattice may be first order.

Finally, for $N\ge 6$, the results in \cite{2013PhRvL.111m7202B} as well as $1/N$ expansions (see e.g. \cite{2008PhRvB..78u4418M}) seem to suggest $\Delta_1 > 1.08$. Then our necessary condition implies that charge two monopole operators can be irrelevant and the phase transition on the rectangular lattice can be second order in agreement with the claims in the literature.

\subsection{Anisotropic deformations in critical $O(n)$ models}\label{ani}

Critical $O(n)$ vector models are canonical examples of conformal fixed points naturally realized in the WLG effective field theory. Under the RG flow, the $O(n)$ critical point is achieved by adjusting one  parameter that is $O(n)$ singlet (e.g. temperature), but the fixed point may be unstable under anisotropic deformations. The stability under anisotropic deformations is an example of emergent symmetry and is subject to our general discussions.

Let us focus on $n=2$ (i.e XY model). It is believed that the $O(2)$ invariant conformal fixed point is unstable under anisotropic deformations with charge $q=2$ and $q=3$, but stable under $q \ge 4$ deformations. Our current best estimate based on the Monte-Carlo (MC) simulation is $\Delta_1 = 0.51905(10)$ and $\Delta_{2}= 1.2361(11)$ \cite{Campostrini:2006ms}. These values are in agreement with the conformal bootstrap analysis of $O(2)$ invariant CFTs in \cite{Kos:2013tga,Kos:2015mba}.
The MC simulations on scaling dimensions of $q=3$ and $q=4$ deformations are also available in \cite{2011PhRvB..84l5136H} as $\Delta_{3} = 2.103(15)$ and $\Delta_{4}= 3.108(6)$.

Now given $\Delta_{1}$ and $\Delta_{2}$, our conformal bootstrap program gives a rigorous upper bound on $\Delta_{3}$. With more precise data to be compared, we have tried to derive the more precise bound by increasing the approximation in our conformal bootstrap program (i.e. $\Lambda = 23$: see Appendix \ref{app:a}). The resulting bound is $\Delta_3 < 2.118$ for $\Delta_1 = 0.51905$ and $\Delta_2 = 1.234$. It turns out that the number quoted above is very close to (or almost saturating) the bound we have obtained.

On the other hand, the estimate of $\Delta_{4} = 3.108(6)$ seems a little below the conformal bootstrap bound $\Delta_{4} < 3.52$. It is not obvious if our bound is saturated but it is at least consistent that our conformal bootstrap bound does not predict that $q=4$ anisotropy must be relevant. 

For $n >2 $, it is a challenging problem to determine exactly when the cubic anisotropy becomes irrelevant. In the large $n$ limit, the cubic anisotropy is believed to be relevant, but for smaller $n$ it is believed to become irrelevant, showing the enhanced $O(n)$ symmetry. The current estimate of the critical $n$ is around $n=3$ (see e.g. \cite{Carmona:1999rm}). However it is still an open question if it is strictly smaller than $3$.

Our conformal bootstrap program might shed some light on this problem. Repeating our analysis now with $O(n)$ symmetry rather than $U(1)$ on the four-point functions $\langle O_{[ij]} O_{[kl]} O_{[mn]} O_{[pq]} \rangle$, we can rigorously show that the cubic anisotropy must be relevant for $n=10$ with no assumptions and for $n=6$ with mild assumptions (e.g. non-conserved vector operators have scaling dimension larger than 3). 
For $n=3$ and $n=4$, the conformal bootstrap bound we have obtained is not conclusive yet.

\section{Discussions}
In this paper, we have numerically proved necessary conditions for emergent symmetry enhancement from discrete symmetry to continuous symmetry under the RG flow. Our conditions are universally valid, but given a concrete model with concrete predictions on critical exponents, we may be able to offer more stringent bounds. Even a modest partial data such as $\Delta_0 = 3-1/\nu$ will make the constraint more non-trivial.  We are delighted to test consistency of future predictions obtained from the other methods upon request.

The symmetry enhancement we have discussed is mainly $U(1)$, but it is possible to discuss non-Abelian enhancement as well from the conformal bootstrap program. For example, there is an interesting conjecture \cite{PhysRevLett.100.137201}\cite{Nahum:2015vka} that the non-compact $CP^1$ model shows further symmetry enhancement to $SO(5)$ by combining N\'eel order parameter and VBS order parameter. However, the conformal bootstrap analysis tells that the currently observed value of $\Delta_1$ (i.e. $\eta_{\mathrm{VBS}} \simeq \eta_{\text{N\'eel}} $) is too small for the conjecture to hold that it has no $SO(5)$ singlet relevant deformation.\footnote{We thank A.~Nahum for discussions. We are informed that D.~Simmons-Duffin had made the same observation. It is tantalizing to note, however, that the bound on the singlet operator in $SO(N)$ symmetric CFTs seems to behave differently at $N=5$ and $6$ (i.e. not a straight line) above the kink than at the other $N$ (i.e. a straight line).}

Finally, we should stress that our discussions are entirely based on the emergent conformal symmetry. In quantum critical systems, this is more non-trivial than in classical critical systems. While the Lorentz invariant conformal fixed points are typically stable under Lorentz breaking deformations allowed on the lattice, it would be important to understand precisely under which condition, the conformal symmetry emerges.

\section*{Acknowledgement}
We thank S.~Aoki, N.~Kawashima, A.~Nahum and T.~Okubo for discussions.
This work is supported by the World Premier International Research Center Initiative (WPI Initiative), MEXT. T.O. is supported by JSPS Research Fellowships for Young Scientists and the Program for Leading Graduate Schools, MEXT.

\bibliographystyle{apsrev4-1} 
\bibliography{accidental} 

\appendix

\section{Technical details about conformal bootstrap implementation}\label{app:a}

The mathematical basis of the numerical conformal bootstrap program is the crossing symmetry of four-point functions in CFTs. In CFTs, we can expand he four-point functions by using the OPE
\begin{align}
O_i \times O_j \sim  \sum_{O} \lambda_{ijO} O , \cr
O_i \times O_j^\dagger \sim \sum_{O} \lambda_{i\bar{j}O} O \ , 
\end{align}
where the assumption of unitarity dictates a reality condition on the OPE coefficient: $\lambda_{ijk}^* = \lambda_{\bar{i}\bar{j}\bar{k}}$, 

\begin{widetext}
In our problems, the crossing symmetry of four-point functions $\langle O_i O_{i}^\dagger O_i O_{i}^\dagger \rangle$  $(i=1,2)$ leads to equations
\begin{align}
    \sum_{O \in O_i \times O^\dagger_i} | \lambda_{i\bar{i}O}|^2 F^{(-)ii,ii} &=0,\cr
    \sum_{O \in O_i \times O_i}|\lambda_{iiO}|^2 F^{(\pm)ii,ii} \pm
    \sum_{O \in O_i \times O^\dagger_i}| \lambda_{i\bar{i}O}|^2 (-1)^{l_O} F^{(\pm)ii,ii}  &= 0 ,
\end{align}
where $l_O$ is the spin of operator $O$ and $F^{(\mp) ij,kl}:= |1-z|^{\Delta_j+\Delta_k} g^{\Delta_{i}-\Delta_j,\Delta_{k}-\Delta_l}_{\Delta,l}(z,\bar{z})\mp \left\{ z\leftrightarrow (1-z) \right\}$ with the conformal block $g^{\Delta_{i}-\Delta_j,\Delta_{k}-\Delta_l}_{\Delta,l}(z,\bar{z})$ normalized as in \cite{Hogervorst:2013kva}. The general framework of conformal bootstrap equations with global symmetry first appeared in \cite{Rattazzi:2010yc} by developing the seminal idea of numerical conformal bootstrap \cite{Rattazzi:2008pe}.

In addition, the crossing symmetry of the four-point function  $\langle O_{1} O_{1}^\dagger O_{2} O_{2}^\dagger \rangle$ gives 
\begin{align}
  \sum _{O \in O_{1} \times O_{2} } |\lambda_{12 O}|^2 F^{(\mp)12,21} &\pm \sum _{O \in O_{1} \times O_{1}^{\dagger} } \lambda_{1\bar{1}O}^* \lambda_{2\bar{2}O}(-1)^{l_O}F^{(\mp)11,22} = 0\cr
    \sum _{O \in O_{1} \times O_{2} } |\lambda_{12 O}|^2 (-1)^{l_O} F^{(\mp)21,21} &\pm \sum _{O \in O_{1} \times O_{2}^{\dagger} } (-1)^{l_O} |\lambda_{12 O}|^2 F^{(\mp)21,21} = 0\cr 
    \sum _{O \in O_{1} \times O_{2}^\dagger } |\lambda_{1\bar{2} O}|^2 F^{(\mp)12,21} &\pm \sum _{O \in O_{1} \times O_{1}^{\dagger} } \lambda_{1\bar{1}O}^* \lambda_{2\bar{2}O} F^{(\mp)11,22} = 0 
\end{align}
\end{widetext}
together with their complex conjugate.

The assumption of unitarity enables us to derive a semi-definite problem from these crossing equations. Once we calculate a table for derivatives of conformal block and its rational function approximation, e.g. by the ordinary differential equations found in \cite{Hogervorst:2013kva} and the recursion relation in \cite{Kos:2014bka}, we can numerically study the semi-definite problem by available optimization programs such as \verb+SDPB+\cite{Simmons-Duffin:2015qma}. We have implemented a code to export the conformal bootstrap problem to \verb+SDPB+ by using a free open-source mathematics software SageMath\cite{sage}\cite{Cboot}.

In the numerical conformal bootstrap program, we introduce a truncation on the search space for the functional whose existence makes the crossing equations impossible to hold. The number of poles in deriving the rational-approximation of the conformal blocks is dictated by the cutoff $\nu_{\mathrm{max}}$ and  the included spins are dictated by the set $S$ (see e.g. \cite{Simmons-Duffin:2015qma} 
for the precise definition).
These must be taken sufficiently large to stabilize the result.
More importantly, the numerical precision is governed  by number of derivatives $\Lambda$ that we use in translating the crossing equations into a semi-definite problem. Note that increasing $\Lambda$ makes our necessary condition only stronger, so our results are, albeit not necessarily being the strongest, still rigorous. 

For our actual study, we use $\Lambda = 15$ for Fig.\ref{fig:charge_3} with $\Delta_1 < 0.565$, $\Lambda=19$ for Fig.\ref{fig:charge_3} with $0.57 < \Delta_1 < 0.575$, and $\Lambda=23$ for Fig.\ref{fig:charge_2}, Fig.\ref{fig:charge_3} in the range $0.580 <\Delta_1 < 0.585$, and Fig.\ref{fig:epsilon_prime}. We try as large as $\Lambda = 33$ in the discussions of section \ref{ani}.


\section{Summary of scaling dimensions in the literature}\label{app:b}

In this appendix, we present a concise summary of scaling dimensions of operators derived or measured in the literature. We have quoted some of them in the main text.

Let us begin with the XY model. The effective action is given by
\begin{align}
S = \int d^3 x \left( \partial_\mu \bar{\phi} \partial^\mu {\phi} + (m^2-m^2_{\mathrm{cr}}) |\phi|^2 +  g|\phi|^4 \right) \ .
\end{align}
On this model, there have been extensive studies on $\nu$ and $\eta$ (i.e.  $\Delta_{0} = 3-1/\nu$ and $\Delta_{1} = \frac{1}{2} + \frac{\eta}{2} $) in the literature including the MC simulations \cite{Campostrini:2006ms}, resummed perturbation theories ($\overline{\mathrm{MS}}$ and MZM) \cite{ZinnJustin:1999bf} as well as conformal bootstrap analysis \cite{Kos:2013tga,Kos:2015mba}. We also find an estimate of higher charged operators that are summarized in table \ref{table:1}. We point out that there is an 8-sigma discrepancy between the most precise MC simulation of $\nu$ \cite{Campostrini:2006ms}  and the most precise measurement of specific heat in the $\lambda$ transition of liquid helium in zero gravity \cite{Lipa:2003zz}.

\begin{table}[htbp]
\resizebox{9.0cm}{!}{
\begin{tabular}{|c||c|c|c|c|c|}
\hline reference
 & $\Delta_0$ &$\Delta_1$ & $\Delta_{2}$ & $\Delta_{3}$ & $\Delta_{4}$ 
 \\ \hline \hline
$\overline{\mathrm{MS}}$ \cite{Calabrese:2002bm,DePrato:2003yd,Carmona:1999rm}  & 1.503(8)&0.5190(25) &1.234(6) &2.10(2) &$3.114(4) $\\ \hline
MZM \cite{Calabrese:2002bm,DePrato:2003yd,Carmona:1999rm} &1.508(3) &  0.5177(12) & 1.234(18)  & $2.103(15)$ & $3.103(8)$ \\ \hline
MC \cite{2011PhRvB..84l5136H} & 1.5112(3) & 0.51905(10) & $1.2361(11)$  & $2.1085(20) $ & $3.108(6) $ \\ \hline

\end{tabular}
}
\caption{Scaling dimensions of operators in the XY model ($N=1$ limit of $CP^{N-1}$ model). The values of $\Delta_0$ and $\Delta_1$ for resummed perturbation theory are taken from \cite{ZinnJustin:1999bf}.}
\label{table:1}
\end{table}

The non-compact $CP^{N-1}$ model has the effective field theory description by a $U(1)$ gauge theory coupled with $N$ charged scalars with $SU(N)$ flavor symmetry:
\begin{align}
S = \int d^3x &\left( \frac{1}{4e^2} F_{\mu\nu} F^{\mu\nu} + D_\mu  \bar{{\phi}}^I D^\mu {\phi}_I  \right. \cr
  & +  (m^2-m^2_{\mathrm{cr}}) |\phi_I|^2 +  g(|\phi_I|^2)^2 \biggr) \ ,
\end{align}
where $I = 1 \cdots N$.  On these models, there has been no compelling theoretical argument except for the large $N$ expansion, whether the system really shows  a critical conformal behavior for small number of $N$. Indeed, there have been several reports that the MC simulation of the quantum spin systems for $N=2$ show the first order phase transition rather than the second order phase transition (e.g. \cite{2008JSMTE..02..009J,2008arXiv0805.4334K}). On the other hand, the scaling behavior of the system, including the ones for the VBS order parameter (i.e. monopole operator) has been studied in various literature. The MC simulation is mainly done on the so-called JQ model and cubic-dimer model (CDM).
We summarize the recent estimate taken from the literature in table \ref{table:2}. $\Delta_{q} > 3$ in the table implies that they assume the phase transition is second order even if the lattice symmetry cannot forbid the symmetry breaking operator $O_{q}$ while $\Delta_{q}<3$ implies that it is first order.

\begin{table}[htbp]
\resizebox{8.8cm}{!}{
\begin{tabular}{|c||c|c|c|c|c|}
\hline reference
 &$\Delta_0$ &$\Delta_1$ & $\Delta_{2}$ & $\Delta_{3}$ & $\Delta_{4}$ 
 \\ \hline \hline
\multicolumn{1}{|c||}
{JQ \cite{2008JSMTE..02..009J,2008arXiv0805.4334K}} & \multicolumn{5}{|c|}{no fixed point} \\ \hline
CDM \cite{2014PhRvB..89a4404S,2015arXiv150402278S} & 1.44(2)  &0.579(8) &1.42(7)&2.80(3) &$>3 $\\ \hline
JQ \cite{2013PhRvL.111h7203P,2015PhRvB..91j4411P}  & 1.15(20) &0.64(4) &  & $>3$ & $> 3$ \\ \hline
JQ \cite{2013PhRvB..88v0408H}& 1.31 &0.68 &  & $>3 $ & $>3 $ \\ \hline
JQ \cite{2013PhRvL.111m7202B}& &$<3$ & $<3$ & $>3 $ & $>3 $ \\ \hline
JQ \cite{2009PhRvB..80r0414L} &1.53(5)  & $0.60(1)$ & $ $ & $ $ & $>3 $ \\ \hline
large $N$ \cite{Dyer:2015zha} &  & $0.63$ & $  1.50 $ & $ 2.55 $ & $3.77 $ \\ \hline

\end{tabular}}
\caption{Scaling dimensions of monopole operators in non-compact $CP^{N-1}$ model ($N=2$).}
\label{table:2}
\end{table}

The $CP^{N-1}$ models with $N\ge3$ have been studied mainly for theoretical interest, but they are regarded as a very good laboratory of the quantum criticality. We summarize the recent estimate of the scaling dimensions taken from the literature in table \ref{table:3} and \ref{table:4}.

\begin{table}[htbp]
\resizebox{8.8cm}{!}{
\begin{tabular}{|c||c|c|c|c|c|}
\hline reference
&$\Delta_0$ &$\Delta_1$ & $\Delta_{2}$ & $\Delta_{3}$ & $\Delta_{4}$ 
 \\ \hline \hline
JQ \cite{2013PhRvB..88v0408H}&1.28 & 0.785 &  2.0 & $>3 $ & $>3 $ \\ \hline
JQ \cite{2013PhRvL.111m7202B}&  &$<3$ & $<3$ & $>3 $ & $>3 $ \\ \hline
JQ \cite{2009PhRvB..80r0414L} &1.46(7) & $0.71(2)$ & $ $ & $ $ & $>3 $ \\ \hline
large $N$ \cite{Dyer:2015zha} &  & $0.755$ & $  1.81 $ & $ 3.10 $ & $4.59 $ \\ \hline
\end{tabular}}
\caption{Scaling dimensions of monopole operators in non-compact $CP^{N-1}$ model ($N=3$).}
\label{table:3}
\end{table}

\begin{table}[htbp]
\resizebox{8.8cm}{!}{
\begin{tabular}{|c||c|c|c|c|c|}
\hline reference &$\Delta_0$
 &$\Delta_1$ & $\Delta_{2}$ & $\Delta_{3}$ & $\Delta_{4}$ 
 \\ \hline \hline
JQ \cite{2013PhRvB..88v0408H} & 1.60 & 0.865 &  & $>3 $ & $>3 $ \\ \hline
JQ \cite{2013PhRvL.111m7202B}& & $<3$  & $>3$ & $>3 $ & $>3 $ \\ \hline
JQ \cite{2009PhRvB..80r0414L}  & 1.57(4) & $0.85(1)$ &  & $ $ & $>3 $ \\ \hline
large $N$ \cite{Dyer:2015zha} &  & $0.880$ & $  2.12 $ & $ 3.64 $ & $5.40 $ \\ \hline
\end{tabular}}
\caption{Scaling dimensions of monopole operators in non-compact $CP^{N-1}$ model ($N=4$).}
\label{table:4}
\end{table}

For $N>4$, the simplest JQ model is not suitable for a study of the N\'eel-VBS transition. In the literature, they have typically added the $J_2$ interaction. We summarize the recent estimate of the scaling dimensions taken from the literature in table \ref{table:5}.

\begin{table}[htbp]
\resizebox{8.8cm}{!}{
\begin{tabular}{|c||c|c|c|c|c|}
\hline reference
&$\Delta_0$ &$\Delta_1$ & $\Delta_{2}$ & $\Delta_{3}$ & $\Delta_{4}$ 
 \\ \hline \hline
JQ (honeycomb) \cite{2013PhRvL.111m7202B}&1.46 & 1.0(1) &    & $>3 $ & $>3 $ \\ \hline
JQ (rectangular) \cite{2013PhRvL.111m7202B}& 1.15& 0.85(10) &  $>3$  & $>3 $ & $>3 $ \\ \hline
JQ \cite{2012PhRvL.108m7201K} & & $1.0(1)$ &  &  & $>3 $ \\ \hline
large $N$ \cite{Dyer:2015zha} &  & $1.00$ & $  2.43 $ & $ 4.18 $ & $6.21 $ \\ \hline
\end{tabular}}
\caption{Scaling dimensions of monopole operators in non-compact $CP^{N-1}$ model ($N=5$).}
\label{table:5}
\end{table} 

In the large $N$ limit, one may directly solve the non-compact $CP^{N-1}$ model. In \cite{1990NuPhB.344..557M,2008PhRvB..78u4418M,Dyer:2015zha}, we find the large $N$ evaluation of the scaling dimensions of charge $q$ monopole operators. For charge one monopoloe operator, we have 
\begin{align}
\Delta_{1} = 0.1246 N + 0.3815 + O(1/N) \ .
\end{align}
They have also computed the scaling dimensions of higher charged monopole operators, which are shown in the above tables in this appendix.  
The $\Delta_0$ was also computed \cite{PhysRevLett.32.292} in the leading $1/N$ expansion as
\begin{align}
\Delta_0 = 2 - \frac{48}{\pi^2 N} + O(1/N^2)  \ .
\end{align}
However, the $1/N$ expansion on this exponent seems less reliable, so we have not shown the values in the tables.

We note that the scaling dimensions of monopole operators have been studied also in $U(1)$ gauge theory coupled with charged {\it fermions} in $2+1$ dimensions. The study is relevant for a certain algebraic spin liquid with possible concrete realizations in nature (e.g. Herbertsmithite) \cite{PhysRevB.72.104404,PhysRevB.77.224413}. The recent analysis of the monopole operators in the fermionic case includes the one from $1/N$ expansions \cite{Pufu:2013vpa}, $\epsilon$ expansions \cite{Chester:2015wao} as well as the conformal bootstrap analysis \cite{Chester:2016wrc}.

Finally, we summarize the properties of so-called collinear fixed point of $O(4) \times U(1)$ invariant WLG theory that is relevant for the QCD chiral phase transition. The effective action is
\begin{align}
S = \int d^3x &\biggl( \mathrm{Tr} \left( \partial_\mu \Phi^\dagger \partial^\mu \Phi \right) + (m^2-m_{\mathrm{cr}}^2) \mathrm{Tr} \Phi^\dagger \Phi \cr
 &+ g_1 (\mathrm{Tr} \Phi^\dagger \Phi)^2 + g_2 \mathrm{Tr} (\Phi^\dagger \Phi)^2  \biggr) \ .
\end{align}
Note that the matrix valued field $\Phi$ is not charged under the vector-like microscopic $U(1)_V$ baryon symmetry, so when we talk about the effective WLG model throughout our work, we always ignore the $U(1)_V$ baryon symmetry.

In table \ref{table:QCD}, which uses the $O(m)\times O(n)$ notation for the flavor symmetry, the (conventionally normalized) $U(1)_A = O(n=2)$ charge $q=4$ operator $\det \Phi \sim \epsilon^{ij} \epsilon^{\bar{i}\bar{j}} \bar{\psi}_{\bar{i}} \psi_{i} \bar{\psi}_{\bar{j}} \psi_j$ we are interested in corresponds to ST sector. To avoid a possible confusion, we note that the $\mathbb{Z}_2$ symmetry discussed in the main text acts on $\det \Phi$ as $-1$, so we used the notation $\Delta_{1} = \Delta_{\mathrm{ST}}$ (despite the conventional $U(1)_A$ assignment). In particular, we did not talk about {\it another} $\mathbb{Z}_2$ symmetry $\Phi \to - \Phi$ that is microscopically non-anomalous and we assume in the effective WLG action. This $\mathbb{Z}_2$ symmetry has nothing to do with the non-trivial $\mathbb{Z}_2$ symmetry proposed in \cite{Aoki:2012yj,Aoki:2013zfa,Kanazawa:2015xna}, which we have discussed in the main text.
\begin{table}[htbp]
\resizebox{8.8cm}{!}{
\begin{tabular}{|c||c|c|c|c|c|c|}\hline
 &$\Delta_\Phi$ & $\Delta_\mathrm{SS}$ & $\Delta_\mathrm{ST}$ & $\Delta_\mathrm{TS}$ & $\Delta_\mathrm{TT}$ & $\Delta_\mathrm{AA}$ \\ \hline \hline
bootstrap &0.558(4)&1.52(5)&0.82(2)&1.045(3)&1.26(1)&1.70(6)\\ \hline
$\overline{\mathrm{MS}}$ &0.56(3)&1.68(17)&1.0(3)&1.10(15)&1.35(10)&1.9(1)\\ \hline
MZM& 0.56(1)& 1.59(14) &0.95(15)&1.25(10)&1.34(5)&1.90(15)\\ \hline
\end{tabular}}
\caption{The scaling dimensions of operators for the $O(4)\times O(2)$ collinear fixed point from \cite{Nakayama:2014sba,Pelissetto:2013hqa,Calabrese:2004nt,Calabrese:2004at}.}
\label{table:QCD}
\end{table}

\section{Bounds on $\Delta_0 '$}\label{app:2.5}
In this appendix, we show a necessary condition on $\Delta_0$ for a unitary CFT to contain only one relevant scalar operator that is neutral under any global symmetries. Let $O_0$ be the lowest such operator with its OPE having the form
\begin{align*}
    O_0 \times O_0 \sim O_0 + O_0 ' + \cdots,
\end{align*}
where $O_0'$ is the other scalar operator with the second lowest scaling dimension $\Delta_0'$.

\begin{figure}[h]
    \begin{center}
  \includegraphics[width=9cm,clip]{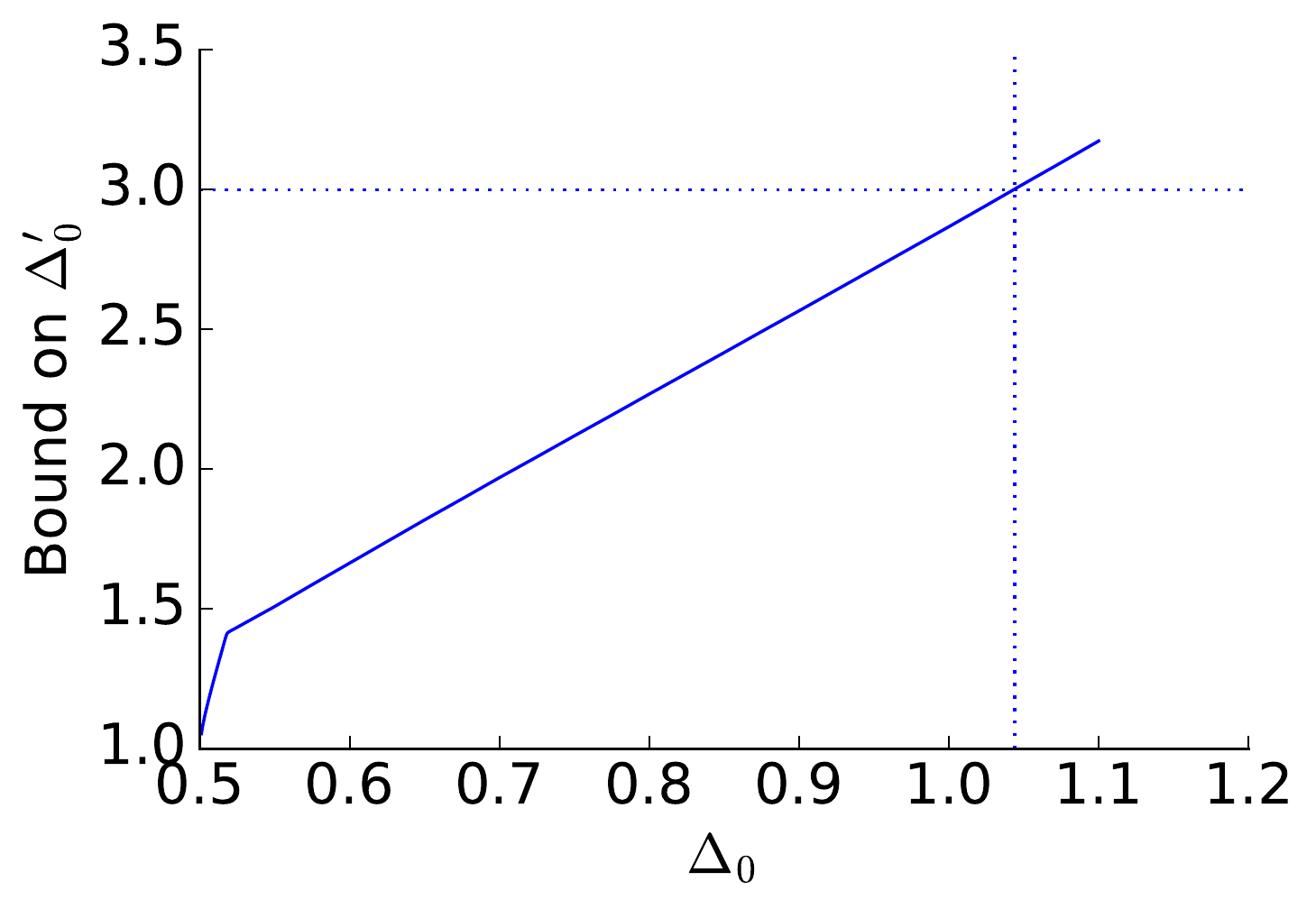}
  \caption{Bounds on the scaling dimensions of the second-lowest neutral scalar operator as a function of $\Delta_0$.}
  \label{fig:epsilon_prime} 
  \end{center}
\end{figure}

We can study the consistency of the four-point function $\langle O_0 O_0 O_0 O_0\rangle$ with the crossing symmetry by using the numerical conformal bootstrap program to derive the upper bound on $\Delta_ 0 '$. 
Compared to the $\mathbb{Z}_2$-odd scalar four-point function studied in \cite{ElShowk:2012ht}\cite{El-Showk:2014dwa}, the only difference is that we have to additionally require the non-negativity of the linear functional acting on the conformal block coming from $O_0$ itself. Although the bounds on $\Delta_0'$ with $O_0$ appearing in the OPE could be weaker than those without $O_0$, we found that these two bounds actually coincide (Fig. \ref{fig:epsilon_prime}). From this plot we obtain the necessary condition $\Delta _0 > 1.044$ (equivalently, $\nu > 0.511$) for the CFTs that contain no singlet relevant scalar operator other than $O_0$ so that the corresponding critical point is achieved by tuning only one parameter.

\section{Supplementary plots}\label{app:3}
\begin{figure}[h]
    \begin{center}
  \includegraphics[width=9cm,clip]{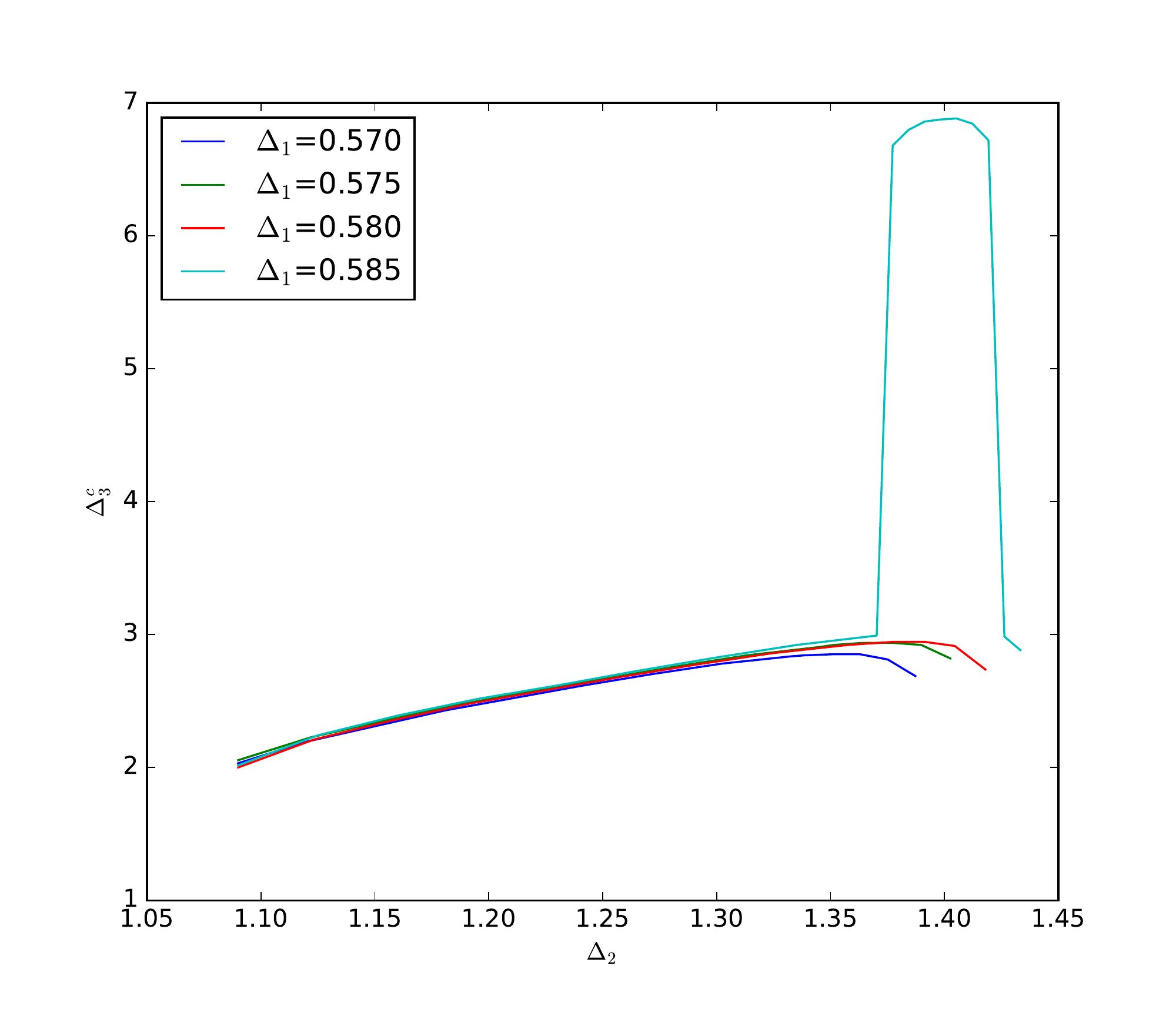}
  \caption{The two-dimensional projection of Fig.\ref{fig:charge_3} in the range $0.57 \le \Delta_1 \le 0.585$. }
  \label{fig:charge_3_3} 
  \end{center}
\end{figure}

In this appendix, we present two-dimensional projected plots of the upper bounds on the scaling dimension of the lowest dimensional charge three scalar operator appearing in $O_1 \times O_2$ OPE as a function of $\Delta_2$ (for a fixed $\Delta_1$). The three-dimensional plot was presented in Fig.\ref{fig:charge_3} of the main text. For the range of $\Delta_{2}$, we recall that $1.08<\Delta_2 < \Delta^c_2 (\Delta_1)$ must hold from the assumption that all the charge four operators are irrelevant and the bound in Fig.\ref{fig:charge_2}.
When $\Delta_1 = 0.585$, the bound shows a jump  as shown in Fig.\ref{fig:charge_3_3}.

A similar phenomenon is observed in the fermionic conformal bootstrap analysis \cite{Iliesiu:2015qra}. It will be interesting to understand why the scaling dimension 3, corresponding to the marginal value of the deformations, plays a special role in the mixed correlator conformal bootstrap program.

\onecolumngrid

\begin{figure}[h]
    \begin{center}
  \includegraphics[width=9cm,clip]{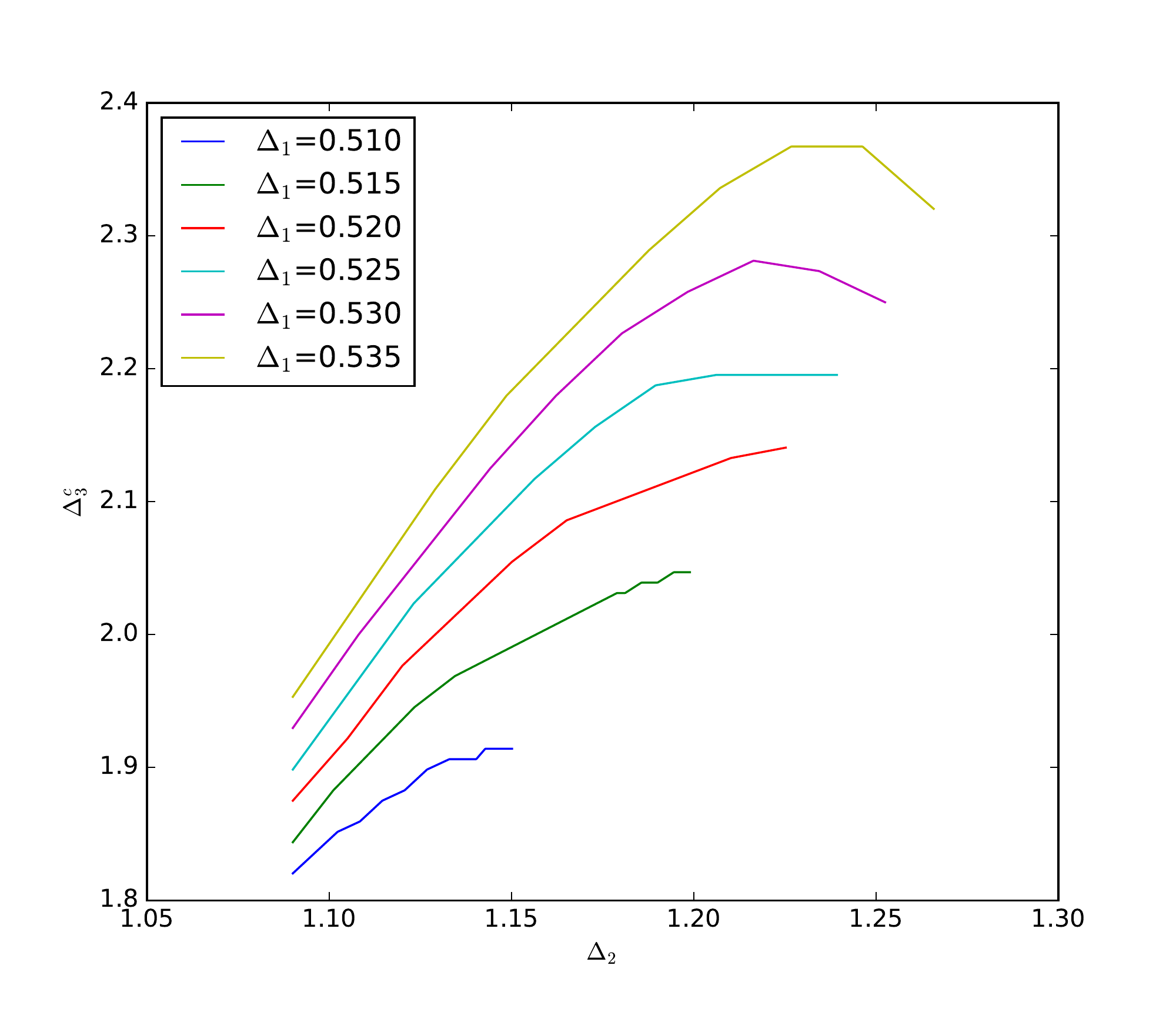}
  \caption{The two-dimensional projection of Fig.\ref{fig:charge_3} in the range $\Delta_1 \le 0.535$. }
  \label{fig:charge_3_1} 
  \end{center}
\end{figure}
\begin{figure}[h]
    \begin{center}
  \includegraphics[width=9cm,clip]{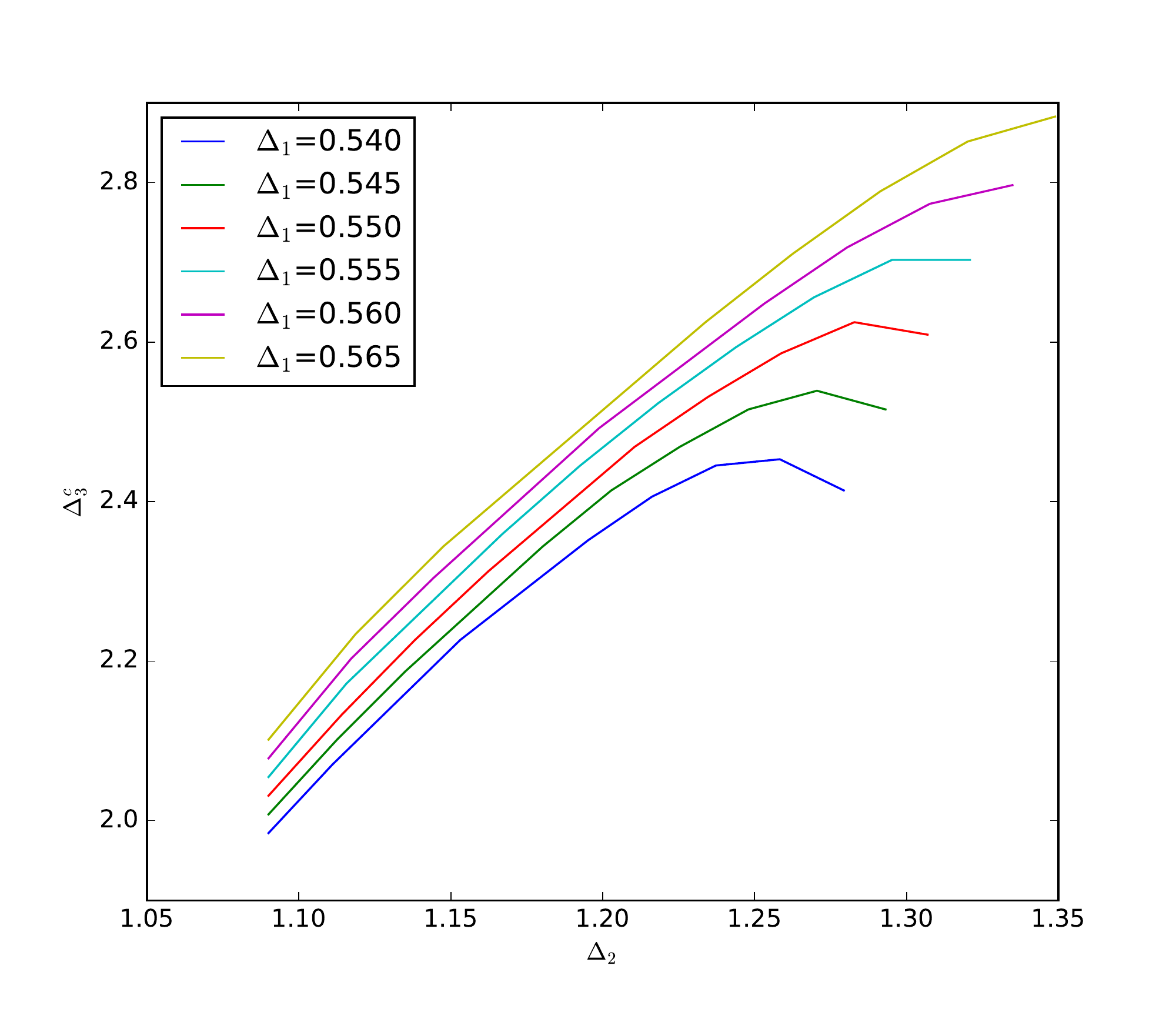}
  \caption{The two-dimensional projection of Fig.\ref{fig:charge_3} in the range $ 0.540 \le \Delta_1 \le 0.565$. }
  \label{fig:charge_3_2} 
  \end{center}
\end{figure}

\twocolumngrid

\end{document}